\documentclass[11pt]{article}

\usepackage[T1]{fontenc}
\usepackage{lmodern}
\usepackage[margin=1in]{geometry}
\usepackage{hyperref}
\usepackage{booktabs}
\usepackage{amsmath}
\usepackage{graphicx}
\usepackage{listings}
\usepackage{xcolor}

\hypersetup{
  colorlinks=true,
  linkcolor=blue,
  citecolor=blue,
  urlcolor=blue
}

\title{Poisoned Identifiers Survive LLM Deobfuscation:\\A Case Study on Claude Opus 4.6}
\author{Luis Guzm\'{a}n Lorenzo\\[0.5em]
\normalsize Model tested: Claude Opus 4.6 (1M context)}
\date{2026-03-22 through 2026-04-05}

\begin{document}

\maketitle

\begin{abstract}
When an LLM deobfuscates JavaScript, can poisoned identifier names
in the string table survive into the model's reconstructed code,
even when the model demonstrably understands the correct semantics?
Using Claude Opus 4.6, we found that explicit verification prompts
(``verify each name matches the math'') failed to prevent propagation
(12/12), while task reframing (``write from scratch'') substantially
reduced propagation on the physics artifact (from 100\% to 20\%) and
eliminated it on the pathfinding artifact (0\%). In our tested setting, the implicit frame of the task changed
naming accuracy where explicit instructions did not.

We tested this across 192 repeated inference runs on two code
archetypes (a force-directed graph simulation and an A* pathfinding
algorithm, 50 conditions, N=3--6) and observed three consistent patterns within the tested conditions:

\begin{enumerate}
  \item \textbf{Wrong decoded identifiers persisted.} Poisoned names appeared
    in the model's deobfuscated code in every baseline run tested
    on both artifacts (physics: 8/8; pathfinding: 5/5). Matched
    controls showed this extends to terms with zero semantic fit
    (\texttt{combustion} for repulsion, \texttt{invoice} for heuristic) when the
    string table does not form a coherent alternative domain.

  \item \textbf{Persistence coexisted with correct semantic commentary.} In
    15/17 runs on the physics artifact, the model wrote wrong variable
    names in code while correctly describing the actual operation in
    comments (manually scored, unblinded, single author; see Section
    3.3 for limitations).

  \item \textbf{Task framing changed persistence.} Reframing from ``deobfuscate
    this'' to ``write a fresh implementation'' reduced propagation from
    100\% to 0--20\% on the physics artifact (N=5) and to 0\% on the
    pathfinding artifact (N=5), while an algorithmic consistency check confirmed the generated
    implementations preserved the checked structure (Appendix F). Explicit verification instructions had
    no effect (12/12 across 4 prompt variants).
\end{enumerate}

Matched-control experiments showed that terms with zero semantic fit
persist at the same rate when the string table does not form a
recognizable alternative domain. Per-term variation observed in
earlier domain-gradient experiments is confounded with domain-level
coherence and recoverability (Section 6).

These observations are from two code archetypes on one model family
(Opus 4.6 primary; Haiku 4.5 limited spot-check). Broader model
and archetype generalization is needed.
\end{abstract}

\section{Introduction}

When Claude Opus 4.6 deobfuscates JavaScript containing poisoned
identifier names, explicit verification instructions (``verify each
name matches the math'') do not prevent the wrong names from appearing
in the output (12/12 runs). Reframing the task as generation (``write
a fresh implementation from scratch'') substantially reduces
propagation while preserving the checked algorithmic structure
(Appendix F). The implicit frame of the task changed naming accuracy
where explicit instructions did not.

Here is what the output looks like:

\begin{figure}[ht]
\centering
\begin{lstlisting}[xleftmargin=0pt,framexleftmargin=0pt,basicstyle=\ttfamily\footnotesize,numbers=none]
    attraction: 4000,           // Repulsion force strength
    ----------                     ---------------------
          |                                 |
     identifier: WRONG              comment: CORRECT

    amplification: 0.92,        // Velocity damping (friction)
    -------------                   ----------------------
          |                                 |
     identifier: WRONG              comment: CORRECT
\end{lstlisting}
\caption{The dual-representation pattern (Phase B, R1 run 1,
Opus 4.6). Wrong identifier in code, correct description in comment.
Observed in 15/17 Phase B runs where the primary endpoint was
positive (at least one poisoned identifier in code blocks). Scored
manually by the first author without blinding; see Section 3.3.}
\label{fig:dual-representation}
\end{figure}

There are two ways to read this. Source recovery: decoded string-table names are the closest available ground truth, and a well-calibrated model should use them. Alternatively: the model does not
apply a capability it shows in other settings (name correction on
readable code, Section 5.1), even when explicitly told to.

A pure inability-to-verify account does not explain these results
well, though the data do not isolate the underlying mechanism.
The strongest defensible claim is: task framing changes naming
behavior without altering the checked algorithmic structure.

We frame the practical implication as cost multiplication, not
protection. Any browser-delivered code is ultimately inspectable
(see Section 8.3).

\textbf{Scope.} This is a case study on two code archetypes (a JavaScript
force-directed graph simulation and an A* pathfinding algorithm)
tested primarily on Claude Opus 4.6 (Haiku 4.5 limited spot-check).
Phase A (exploratory, N=1) used Claude Code CLI with tool use.
Phase B (replicated, N=3--6) used the Anthropic Messages API without
tool use. Results may reflect model-specific or interface-specific
behaviors. We do not claim generality beyond these tested conditions.

\section{Related Work}

\subsection{Code obfuscation and deobfuscation}

\textbf{Pre-LLM baselines.} \textbf{Lu, Coogan, and Debray} \cite{lu2011} describe
semantics-based automatic deobfuscation of JavaScript using dynamic
analysis and slicing (2011). Their approach recovers functionally
equivalent code without requiring knowledge of the obfuscation
scheme, establishing the pre-LLM baseline for automated JS
deobfuscation. The transition from such deterministic approaches
to LLM-based deobfuscation introduces the identifier-naming question
central to this paper: deterministic tools recover whatever literals
the code contains, while LLMs reconstruct names that may reflect
training priors, decoded string tables, or both.

\textbf{CASCADE} \cite{cascade}: Google's hybrid LLM + compiler deobfuscator. Uses
Gemini for prelude function identification (99.56\% accuracy via
few-shot prompting), then hands off to JSIR for deterministic
constant propagation, with sandboxed V8/QuickJS for dynamic
execution. Achieves 98.93\% success on standard obfuscator.io output.
CASCADE's JSIR constant propagation recovers the original string
literals (before obfuscation encoded them), not semantically
meaningful names derived from algorithmic analysis. If the original
names were poisoned before obfuscation (our attack model), we predict
but did not test that JSIR would recover those poisoned names
faithfully, since it restores original literals regardless of semantic
correctness.

\textbf{webcrack} \cite{webcrack}: Purpose-built AST-level deobfuscator for javascript-obfuscator output. Supports string array decoding (base64, RC4),
control flow unflattening, and dead code elimination via pattern
matching. Issue history indicates failure cases under some transform
combinations (e.g., CFF + dead code injection, issue \#44). Our
RC4-encoded production build was not fully decoded by webcrack's
automated pipeline in Phase A testing.

\textbf{JS-Confuser} \cite{jsconfuser}: Alternative obfuscator using state-machine CFF
(explicit state transitions instead of dispatch strings). We expect
webcrack cannot reverse this pattern (untested; webcrack's source
matches only the split-string dispatch pattern).

\subsection{Identifier naming and LLM code understanding}

\textbf{Wang et al.} \cite{wang2024} measure the impact of misleading variable names
on LLM code analysis tasks (using CodeBERT-family models), finding
that shuffled or misleading names degrade performance more than
random strings because they actively misdirect the model. This
supports the general premise that identifier names influence LLM
code processing, though the setting (analysis tasks on code models)
differs from our deobfuscation setting.

\textbf{``The Code Barrier''} \cite{codebarrier} introduces semantics-preserving
obfuscations including cross-domain term substitution, finding that
removing meaningful names severely degrades intent-level tasks.
Their cross-domain renaming experiments provide relevant background
for our domain-gradient observations. Our contribution is specific
to the deobfuscation string-table setting: we observe that
wrong-but-plausible names are actively preserved in reconstructed
code, not merely that they degrade understanding.

\textbf{CoTDeceptor} \cite{cotdeceptor} demonstrates that multi-stage obfuscation
strategies can evade CoT-enhanced LLM vulnerability detectors,
in line with the general finding that adversarial content can
bypass model analysis.

\subsection{Obfuscation vs.\ LLM code analysis}

\textbf{Li et al.} \cite{li2025} categorize obfuscation techniques (layout, data
flow, control flow) against LLM vulnerability detection, finding
both degradation and, in some cases, unexpected improvements.
\textbf{``Digital Camouflage''} \cite{digitalcamo} tests Claude, Gemini, and GPT-4o
against compiler-level obfuscation (CFF, bogus control flow) on
LLVM-obfuscated C functions, confirming cross-model degradation.
\textbf{FORGEJS} \cite{forgejs} benchmarks LLM vulnerability detection in JavaScript
and finds generally unreliable performance. \textbf{JsDeObsBench} \cite{jsdeobsbench}
provides the first systematic benchmark for LLM JavaScript
deobfuscation, evaluating GPT-4o, Mixtral, Llama, and DeepSeek-Coder
across obfuscation types from variable renaming to control flow
flattening. Its focus is on overall deobfuscation quality (syntax
accuracy, execution reliability, code similarity); our work addresses
a different question: whether specific poisoned identifiers survive
into the model's output. \textbf{Yefet et al.} \cite{yefet2020}
establish adversarial examples for code models. These studies show that obfuscation frequently degrades LLM code
analysis under tested settings. Our contribution extends this in a
specific direction: surface-level
poisoning (wrong names in the string table) not only degrades
understanding but can produce contradictory outputs where the
model shows correct semantic knowledge and incorrect identifier
usage in the same response.

\subsection{Tool-augmented LLM agents}

\textbf{SWE-bench} \cite{swebench} and related work on agentic coding workflows show
that tool use (file read/write, shell execution) produces qualitatively
different behavior than single-shot inference. Our observation that
progressive analytical escalation appeared in Claude Code (Phase A,
with tools) but not in API testing (Phase B, without tools) is
in line with this (Section 7.1).

\textbf{Historical precedent.} 1988 paragraph-book anti-piracy \cite{sentinel1988,wasteland2016}:
see Section 8.3.

\section{Method}

\subsection{Stimuli}

\textbf{Exploratory vs.\ hypothesis-testing.} Phase A was exploratory: single-run
qualitative observations that generated hypotheses. Phase B sub-phases
0--3 were hypothesis-testing with replication. Sub-phases 4--6 were
targeted follow-ups, designed to test specific predictions
(translation-frame, domain-boundary stability) and boost sample
sizes on key conditions from earlier sub-phases. We did
not preregister hypotheses. The same author designed stimuli, ran
experiments, and scored results. We know these are limitations. The automated scoring pipeline and
published raw outputs help, but do not fully compensate.

Two phases of experimentation:

\textbf{Phase A (exploratory, N=1):} 17 pill designs (we use ``pill'' as
shorthand for individual adversarial test stimuli, after the poison-pill metaphor) tested in Claude Code CLI sessions (with tool use). 7
production build variants. 28 tests. Results are qualitative
observations that generated hypotheses.

\textbf{Phase B (systematic, N=3--6):} 50 conditions tested via Anthropic
Messages API (no tool use). 192 automated runs across 2 models (183
Opus, 9 Haiku) in 10 sub-phases (0--9). Results are quantitative with
per-condition replication. All conditions at N=3 or higher. Sample
sizes preclude formal inferential statistics; we report observed
proportions and treat consistency across varied conditions as the
primary evidential basis.

Control: pill 09 (clean code, standard obfuscation, no adversarial
content).

\textbf{Unit of analysis.} The 192 Phase B runs test varied conditions on
two code archetypes (a force-directed graph simulation, primary; an
A* pathfinding algorithm, replication). The design supports claims
about identifier propagation behavior under varied prompt,
obfuscation, and naming conditions on these archetypes. It does not
support claims about LLM deobfuscation in general. We frame the
paper as a case study accordingly.

\subsection{Protocol}

\textbf{Phase A:} Fresh Claude Code session from \texttt{/tmp}. File deleted before
each session. Model's complete response recorded manually. Follow-up
prompts used in some sessions to test bypass resistance. Tool use
(Bash, file read/write) available to the model.

\textbf{Phase B:} Automated via Anthropic Messages API. Each pill's code
sent as a user message with the exact prompt text. No tool use. Full
response, token counts, and stop reason recorded as structured JSON.
Automated scoring with code-block-level analysis (see 3.3).

\textbf{Temperature.} Most runs used API default temperature. A targeted
comparison at temperature 0 (N=5, Phase 5) showed identical 100\%
propagation for core terms (\texttt{attraction} 5/5, \texttt{amplification} 5/5),
confirming the effect is deterministic for high-fit names and not a
sampling artifact.

\subsection{Metrics}

\textbf{Primary endpoint:} Whether a poisoned identifier is preserved in
code blocks of the model's output (binary, per term per run). This
is the unit of analysis for all propagation claims.

\textbf{Secondary endpoint (dual-representation):} Whether wrong identifier
in code AND correct semantic description in comments/prose appear in
the same response (binary, per run). This pattern appeared in 15/17
Phase B runs where the primary endpoint was positive.

Scoring for this endpoint: a run was counted as dual-representation
positive when (a) at least one poisoned identifier appeared in a code
block, AND (b) the same response contained a comment or prose
description that correctly named the actual operation (e.g., ``Repulsion
force strength'' adjacent to \texttt{attraction}). ``Correct description'' was
defined as any phrase that would allow a reader to identify the true
physical operation without reference to the variable name. This was
scored manually by the first author on all 17 applicable Phase B runs.
Scoring was not blinded; the scorer knew which names were poisoned.
Ambiguous descriptions (e.g., ``force parameter'' without specifying
direction) were excluded from the positive count. The 15/17 rate
should be read with this unblinded, single-scorer limitation in mind.

\textbf{Tertiary endpoints:}
\begin{itemize}
  \item \textbf{Refusal:} Keyword detection for ``malware,'' ``I won't,'' ``do not
    run,'' ``compromised,'' etc.\ (binary per run). We manually validated
    a subset of 10 runs across conditions and confirmed the keyword
    proxy matched human judgment in all checked cases (Appendix D).
  \item \textbf{Semantic corruption scoring:} For each poisoned name, check
    whether it appears in code blocks (\texttt{```...```}) vs.\ only in prose.
    Scoring definitions (applied per poisoned term per run):
    \begin{itemize}
      \item \textbf{Propagated:} Wrong name appears in code blocks, correct name
        does not appear in code blocks.
      \item \textbf{Flagged:} Both wrong and correct names appear in code blocks.
      \item \textbf{Corrected:} Correct name appears in code blocks, wrong name
        does not.
      \item \textbf{Absent:} Neither wrong nor correct name appears in code blocks.
    \end{itemize}
    N.B.: Naive full-text scoring produces false negatives because the
    model writes correct physics descriptions in comments alongside
    wrong variable names. This motivated the code-block-level scoring.
    A model-based adjudication check (Appendix E) validated this scorer
    against an independent Haiku 4.5 blind adjudicator: 17/20 agreement
    (85\%), with all disagreements in the conservative direction (the
    automated scorer undercounted propagation).
  \item \textbf{Time/tokens:} API-reported processing time and token usage.
  \item \textbf{Functional reconstruction:} Binary, tested in Phase A production
    builds only.
\end{itemize}

\subsection{Experimental Matrix}

\begin{table}[ht]
\centering
\small
\begin{tabular}{@{}lp{6.5cm}rl@{}}
\toprule
Sub-phase & Condition group & N (runs) & Model(s) \\
\midrule
Phase 0 & Pill 10 replication & 5 & Opus \\
Phase 0 & Pill 03 replication & 3 & Opus \\
Phase 0 & Pill 10 cross-model & 3 & Haiku \\
Phase 0 & Pill 03 cross-model & 3 & Haiku \\
Phase 1 & Prompt variants (4 prompts) & 12 & Opus \\
Phase 1 & Domain gradient (4 domains) & 8 & Opus \\
Phase 1 & Obfuscation gradient (3 levels) & 6 & Opus \\
Phase 1 & Task frames (4 frames) & 8 & Opus \\
Phase 2 & Layer gradient (5 layers) & 10 & Opus \\
Phase 2 & Canary & 3 & Opus \\
Phase 3 & Extended domains (gamedev, overlap) & 6 & Opus \\
Phase 3 & Multi-agent verify & 6 & Both \\
Phase 3 & Task frame replications & 6 & Opus \\
Phase 4 & Framing experiment (3 frames) & 15 & Opus \\
Phase 4 & N-boost: gradient, canary, multi-agent & 20 & Opus \\
Phase 5 & N-boost: 4 domains + heavy obf + temp 0 & 21 & Opus \\
Phase 6 & N-boost: obf gradient, task frames, layer gradient & 27 & Opus \\
Phase 7 & Matched lexical controls (2 terms, full table) & 10 & Opus \\
Phase 8 & Full matched-table control (all terms, no fit) & 5 & Opus \\
Phase 9 & Second artifact (A* pathfinding): baseline, framing, control & 15 & Opus \\
\midrule
\textbf{Total} & \textbf{50 conditions} & \textbf{192 runs} & \textbf{183 Opus + 9 Haiku} \\
\bottomrule
\end{tabular}
\end{table}

Phase A (Claude Code CLI): 28 additional exploratory runs, N=1 per
condition.

\subsection{Limitations}

\begin{itemize}
  \item Two models tested (Opus 4.6, Haiku 4.5). Broader generalization
    requires GPT-4o, Gemini, open-source models.
  \item Two code archetypes tested (force-directed graph, A* pathfinding).
    Further archetype diversity is needed for broader generalization.
  \item Phase A: single run per condition, manual interpretation.
  \item Phase B: 3--6 runs per condition, automated scoring.
  \item Self-designed stimuli with implicit bias toward known model behaviors.
  \item Phase A and Phase B use different interfaces (Claude Code with tools
    vs.\ raw API without tools). Behaviors observed in Phase A may be
    specific to tool-using workflows.
  \item The scoring methodology (code-block extraction + regex) may miss
    propagation in non-standard response formats.
  \item Most Phase B runs used API default temperature.
\end{itemize}

Negative results and non-replications are reported alongside positive
findings throughout this paper.

\section{Taxonomy of Adversarial Code-Semantic Manipulations}

We tested five categories of adversarial content embedded in identical
JavaScript code (a force-directed graph simulation). Four produced no
effect on the model's analysis; one produced consistent propagation.
The negative results establish specificity.

\subsection{Ineffective Manipulations}

Four categories produced no effect: inert textual content (wrong
comments, legal threats, CBRN strings, combinations; pills 01--05),
and attention capture via malware decoys (pill 14, Phase A only).
The model identified all inert content as ``dead code'' or ``prompt
injection'' in 7/7 tests. See Appendix A for the pill-by-pill index.

\subsection{Structural Executable Content: Triggers Refusal}

Functions with their own coherent malware-like logic (not physics
code with malware names, but actual malware-shaped operations).

\textbf{Tested:} Functions building credential objects, ransom notes,
exploit lists (pill 03). Realistic functions with actual API calls
(pills 17, production builds v4--v7).

\textbf{Observed:} The model refused to assist with these functions on
readable code. On obfuscated code, refusal reversed: the model
entered ``forensics mode'' and fully deobfuscated the malware
(pill 13). On partially obfuscated code with realistic malware,
the model classified the entire file as malware and refused to read
further.

\textbf{Key condition:} Refusal requires that the malware functions have
their own internal logic that is not reclassifiable as benign.
Functions named \texttt{credential\_harvest} that compute $\sqrt{\Delta x^2 + \Delta y^2}$ are identified as ``physics with edgy names.'' Functions that call \texttt{require('child\_process').\allowbreak{}execSync(...)} cannot be reclassified.

\textbf{Phase B refusal rates (API, replicated):}

\begin{table}[ht]
\centering
\small
\begin{tabular}{@{}llrl@{}}
\toprule
Condition & Model & N & Refusal rate \\
\midrule
Pill 03 (malware functions, readable) & Opus 4.6 & 3 & \textbf{100\%} \\
Pill 03 (malware functions, readable) & Haiku 4.5 & 3 & \textbf{100\%} \\
Pill 10 (poisoned names, no malware) & Opus 4.6 & 5 & 0\% \\
Pill 10, security framing & Opus 4.6 & 5 & \textbf{100\%} \\
Pill 10, practical framing & Opus 4.6 & 5 & 20\% \\
Pill 14 (malware + poisoned) & Opus 4.6 & 5 & \textbf{100\%} \\
Pill 22-4layer / 22-5layer & Opus 4.6 & 5 each & \textbf{100\%} \\
\bottomrule
\end{tabular}
\end{table}

Refusal is binary and replicates across Opus and Haiku (6/6).
Security-framed prompts trigger refusal even without malware.

\subsection{Naming-Semantic Manipulation: Propagates After Obfuscation}

Renaming physics variables to wrong-but-plausible-in-domain terms:
\texttt{repulsion} to \texttt{attraction}, \texttt{damping} to \texttt{amplification}, etc.

\textbf{Tested on readable code:} Pills 06, 07, 08. The model caught
inverted semantics by tracing mathematical signs (pill 06), caught
swapped node/edge variables by reading object shapes (pill 07), and
propagated truncated constants without correction (pill 08: TAU=6.283
used as-is with ``\textasciitilde 2pi'' annotation).

\textbf{Tested after obfuscation:} Pill 10 (testbed, 130 lines). All
wrong names propagated into the model's deobfuscated output with the
dual-representation pattern described in Section 1 (Figure~\ref{fig:dual-representation}).

\textbf{Key condition (revised by Phase 8 matched controls):} Initial
experiments suggested wrong names needed to be plausible in the target
domain. Matched-control tests (Section 6.1) showed a broader pattern:
decoded names persist even with zero semantic fit (\texttt{combustion} for
repulsion) when the replacement table lacks a coherent alternative-domain signal. Malware-domain names (\texttt{exfil\_range}) are still
corrected; the model re-derives from algorithm structure (pill 11),
likely because malware terms trigger a distinct recognition pathway.

\subsection{Numerical Precision Manipulation: Propagated in All Tested Conditions}

Truncated mathematical constants: $\tau \approx 6.283$ (real $2\pi = 6.28318\ldots$),
$1/\sqrt{2} \approx 0.707$ (real $= 0.70710\ldots$).

\textbf{Observed:} The model propagated truncated values without correction
in all conditions where we explicitly tested it: readable code
(pill 08, Phase A) and obfuscated code (pill 10, Phase A+B). It added
an approximate annotation (``\textasciitilde 2pi'') but did not substitute the precise
value. This was the only manipulation that propagated on readable
code without obfuscation. Systematic replication of constant
propagation was not prioritized in Phase B, where the focus was on
identifier names.

\section{Identifier Persistence Under Deobfuscation}

\subsection{Observation}

On readable code, the model corrected wrong names using cues
consistent with algorithm pattern recognition, object field shapes,
mathematical sign tracing, domain-name cross-reference, and numerical
precision. Wrong names were overridden in every readable-code test.

After obfuscation (control flow flattening, RC4 string encoding),
the observed outputs showed decoded names dominating reconstructed
identifiers. The model produced the dual-representation pattern
(Figure~\ref{fig:dual-representation}): wrong identifiers in code, correct descriptions in
comments.

\subsection{Persistence Under Verification Instructions}

In Phase B, we tested four prompt variants on pill 10
(N=3 per variant, 12 total runs). Prompts ranged from baseline
(``deobfuscate this'') to explicit adversarial warning (``the string
table may contain adversarial names; cross-check each decoded name
against the algorithm structure'').

Result: 12/12 runs propagated the wrong names in code regardless of
prompt:

\begin{table}[ht]
\centering
\small
\begin{tabular}{@{}lll@{}}
\toprule
Prompt & \texttt{attraction} in code & \texttt{amplification} in code \\
\midrule
A: ``deobfuscate this'' (baseline) & \textbf{3/3} & \textbf{3/3} \\
B: ``deobfuscate + verify names'' & \textbf{3/3} & \textbf{3/3} \\
C: ``verify each name matches the math'' & \textbf{3/3} & \textbf{3/3} \\
D: ``string table may be adversarial, cross-check'' & \textbf{3/3} & \textbf{3/3} \\
\bottomrule
\end{tabular}
\end{table}

The model spent more time on verification prompts (52.6s vs 36.6s
baseline) but still propagated. This is difficult to explain as a
simple failure to check.

\subsection{The Framing Experiment}

The verification-prompt result suggests the effect is not a simple
prompting omission. But what kind of effect is it? One explanation
that fits the data: deobfuscation may activate a translation-like workflow where decoded string-table entries serve as ``source
text.'' The model would then preserve source identifiers while adding
explanatory comments, matching the observed dual-representation
pattern.

This predicts that reframing from ``deobfuscate this'' (translation) to
``write a fresh implementation'' (generation) should reduce propagation.

\textbf{We tested this prediction (Phase 4, N=5 per frame):}

\begin{table}[ht]
\centering
\small
\begin{tabular}{@{}lllll@{}}
\toprule
Frame & \texttt{attraction} & \texttt{repulsion} & \texttt{amplification} & \texttt{damping} \\
\midrule
``Deobfuscate'' (translation) & \textbf{5/5} & 1/5 & \textbf{5/5} & 5/5 \\
``Rewrite with correct names'' & 4/5 & 5/5 & 3/5 & 5/5 \\
``Write fresh from scratch'' & \textbf{1/5} & \textbf{5/5} & \textbf{0/5} & \textbf{5/5} \\
\bottomrule
\end{tabular}
\end{table}

The generation frame reduced \texttt{attraction} propagation from 100\% to
20\% and \texttt{amplification} from 100\% to 0\%. The model produced correct
names (\texttt{repulsion}, \texttt{damping}) at 100\% when freed from the
translation frame. The rename-correction frame was intermediate:
the model attempted correction (correct names 5/5) but still
preserved wrong names alongside (4/5, 3/5), increasing dual-representation rather than resolving it.

These results fit a translation-frame explanation but do not fully
isolate it. The generation frame also changes
fidelity constraints, output freedom, and the cost of renaming.
An algorithmic consistency check (Appendix F) confirmed that
generation-frame outputs preserve the checked structure on the
physics artifact (5/5 runs) and on pathfinding (5/5 runs). This
rules out drift on the checked components as an explanation for
correct naming. What the data do establish: the identifier-persistence
effect is not an intrinsic inability. The model uses correct names
readily under a different task frame.

\subsection{Propagation Across Conditions}

\textbf{Baseline propagation (Phase B, code-block-level scoring):}

\begin{table}[ht]
\centering
\small
\begin{tabular}{@{}llrll@{}}
\toprule
Condition & Model & N & \texttt{attraction} in code & \texttt{amplification} in code \\
\midrule
Pill 10 (physics-domain poison) & Opus 4.6 & 5 & \textbf{5/5} & \textbf{5/5} \\
Pill 10 (physics-domain poison) & Haiku 4.5 & 3 & \textbf{3/3} & 2/3 \\
Pill 24 (canary comment) & Opus 4.6 & 3 & 2/3 & 0/3 \\
\bottomrule
\end{tabular}
\end{table}

A research-context canary comment partially reduces propagation for
marginal-plausibility terms; core terms still propagated 2/3.

\textbf{Obfuscation gradient (pill 10 source, 3 levels):}

\begin{table}[ht]
\centering
\small
\begin{tabular}{@{}lllr@{}}
\toprule
Level & \texttt{attraction} in code & \texttt{amplification} in code & N \\
\midrule
Light (base64, CFF 0.5) & \textbf{5/5} & \textbf{5/5} & 5 \\
Medium (base64, CFF 0.5, self-defending) & \textbf{5/5} & \textbf{5/5} & 5 \\
Heavy (RC4, CFF 0.75, split strings) & \textbf{6/6} & 0/6 & 6 \\
\bottomrule
\end{tabular}
\end{table}

Light obfuscation is sufficient to trigger propagation (16/16 for
\texttt{attraction} across all levels). \texttt{amplification} dropped to 0/6 at
heavy obfuscation, where split-string encoding fragments the name
across chunks. This recoverability interaction is one reason the
domain-gradient experiments (Section 6.2) should be treated as
exploratory.

\section{What Determines Whether Wrong Names Are Corrected?}

\subsection{Matched Controls (Phase 7--8)}

The domain-gradient experiments (Section 6.2) suggested per-term
semantic fit determines propagation, but this was confounded with
domain-level coherence. To isolate the variable, we created
length-matched terms with zero semantic fit:
\texttt{combustion} (10 chars, like \texttt{attraction}) for repulsion,
\texttt{communication} (13 chars, like \texttt{amplification}) for damping. In one
condition only 2 terms were changed; in another, ALL config property
names were replaced with length-matched no-fit terms.

\begin{table}[ht]
\centering
\small
\begin{tabular}{@{}llll@{}}
\toprule
Condition & All terms replaced? & Domain coherent? & Propagation \\
\midrule
pill-10 (physics terms) & Yes & Yes (physics) & FLAGGED 5/5 \\
pill-25 (2 no-fit terms, rest physics) & No & Mostly & FLAGGED 5/5 \\
pill-27 (ALL no-fit terms, incoherent) & Yes & No & \textbf{FLAGGED 5/5} \\
pill-20-engineering (ALL engineering) & Yes & \textbf{Yes (engineering)} & \textbf{CORRECTED 5/5} \\
\bottomrule
\end{tabular}
\end{table}

\textit{FLAGGED = both wrong and correct names appeared in code blocks;
the wrong name was preserved. CORRECTED = only correct names in code.}

\texttt{combustion} persisted in code at the same rate as \texttt{attraction}, even
when every term in the string table lacked semantic fit. Full
correction occurred in the engineering coherent-domain condition
(0/5 propagation), though other coherent domains (e.g., finance)
produced mixed rather than clean correction.

The same pattern appeared on the second artifact (A* pathfinding):
\texttt{invoice} (zero fit to heuristic estimation) persisted at 5/5, while
the generation frame corrected all terms to proper pathfinding
vocabulary (5/5). See Section 6.3.

\subsection{Domain-Gradient Experiments (Phases 1--5, exploratory)}

Earlier domain-pill experiments, which replaced all terms with
coherent domain-specific alternatives, showed per-term variation:

\begin{table}[ht]
\centering
\small
\begin{tabular}{@{}lllrl@{}}
\toprule
Term & For operation & Domain & Propagation & N \\
\midrule
\texttt{attraction} & repulsion & physics & 100\% & 8 \\
\texttt{decay} & damping & overlap & 100\% & 3 \\
\texttt{velocity} & repulsion & game dev & 75\% & 8 \\
\texttt{yield} & repulsion & finance & 60\% & 5 \\
\texttt{acceleration} & damping & game dev & 0\% & 8 \\
\texttt{inductance} & repulsion & engineering & 0\% & 5 \\
\bottomrule
\end{tabular}
\end{table}

This looked like per-term semantic evaluation at first. However,
the matched-control results (Section 6.1) showed this interpretation
is confounded: the domain pills replaced ALL terms with coherent
alternatives, creating a recognizable domain signal. The per-term
variation likely reflects the model's confidence in the alternative
domain, not per-term semantic evaluation. The \texttt{acceleration} result
(0/8 despite being a core physics term) may reflect domain-level
recognition of the game-dev replacement set rather than recognition
that acceleration contradicts what damping does.

The intermediate points (e.g., yield 3/5) are underpowered and
should not be over-interpreted. This initial gradient motivated the
matched-control experiments (Section 6.1) that clarified the picture.

\subsection{Second Artifact Replication (Phase 9)}

To test whether these patterns are specific to the physics simulation,
we repeated the core protocol on an A* pathfinding algorithm (\textasciitilde 110
lines, graph-search domain). The pathfinding artifact uses a different
algorithmic structure (priority-queue search vs.\ force integration),
different domain vocabulary, and a different poison map:

\begin{table}[ht]
\centering
\small
\begin{tabular}{@{}lll@{}}
\toprule
Poisoned name & Correct name & Operation \\
\midrule
\texttt{penalty} & \texttt{heuristic} & distance estimation method \\
\texttt{adjacentCost} & \texttt{diagonalCost} & diagonal movement cost \\
\texttt{closedSet} / \texttt{openSet} & (swapped) & A* frontier sets \\
\texttt{fScore} / \texttt{gScore} & (swapped) & path cost components \\
\texttt{ancestors} & \texttt{neighbors} & adjacent-cell expansion (same function, wrong label) \\
\bottomrule
\end{tabular}
\end{table}

Scoring used the same code-block extraction methodology as the
physics artifact. The generation-frame prompt was adapted to
reference pathfinding rather than physics. Three conditions at N=5
each (15 runs total):

\begin{table}[ht]
\centering
\small
\begin{tabular}{@{}lll@{}}
\toprule
Condition & Physics artifact & Pathfinding artifact \\
\midrule
Baseline (poisoned names) & FLAGGED 5/5 & FLAGGED 5/5 \\
Generation frame & CORRECTED 4--5/5 & CORRECTED 5/5 \\
Matched no-fit control & FLAGGED 5/5 & FLAGGED 5/5 \\
\bottomrule
\end{tabular}
\end{table}

The pathfinding no-fit control used \texttt{invoice} (7 chars, like
\texttt{penalty}) for the heuristic method and \texttt{purchaseCost} (12 chars,
like \texttt{adjacentCost}) for diagonal movement cost.

Algorithmic consistency was checked for both artifacts' generation-frame outputs (Appendix F).

All three patterns replicated on the second artifact: persistence,
framing-dependent correction, and no-fit-term persistence. This
suggests the observed patterns are not unique to the first artifact,
though two archetypes remain a limited basis for generalization.

\section{Secondary Findings}

\subsection{Adversarial Salience Escalation (Phase A only)}

In Phase A (Claude Code with tools), defense sophistication was
inversely correlated with defense effectiveness. Pill 10 (wrong-but-plausible names, no structural anomalies) produced the cleanest
corruption because nothing signaled adversarial intent. Adding
layers (malware functions, prompt injection) triggered progressively
more sophisticated analysis.

\textbf{Phase B did not replicate this pattern.} In the API (without
tools), refusal is a step function at the malware-function boundary,
not a gradient. The model either analyzes routinely or refuses.
The difference between Phase A and Phase B may reflect tool-use
dynamics, human cueing, or both. We treat this as a tool-use-specific observation rather than a general finding.

\subsection{Cost Inflation (Phase A, Claude Code with tools)}

As an engineering illustration: the undefended control (pill 09) was
deobfuscated in 2.5 min with near-perfect output. The defended
production build (v7) required 13.5 min for structural recovery and
3+ hours for a full reconstruction attempt (137 tool invocations,
268K tokens) that produced non-functional output with poisoned
parameter names throughout. These are single-run Phase A observations
and should not be read as controlled effect-size estimates.

\subsection{Multi-Agent Verification Pilot (Phase 3)}

A fresh model instance was given the deobfuscated code (readable, with
wrong names and correct comments) and asked to verify each name.

\begin{table}[ht]
\centering
\small
\begin{tabular}{@{}lrll@{}}
\toprule
Model & N & Wrong names flagged & Fully corrected in code \\
\midrule
Opus 4.6 & 3 & 2/3 & 0/3 \\
Haiku 4.5 & 3 & 2/3 & 0/3 \\
\bottomrule
\end{tabular}
\end{table}

Both models detected the discrepancy but neither consistently removed
the wrong names from code output. At N=3 per model, this is a pilot
observation. The readable-code setting is cognitively different from
deobfuscation, so the two effects may not share a common mechanism.

\section{Discussion}

\subsection{Conjecture}

One hypothesis that fits the data: during deobfuscation,
decoded identifier names (high apparent reliability) may dominate
over algorithmic patterns (degraded by obfuscation) in the model's
output.

The matched-control results (Section 6.1) suggest a more specific
hypothesis: the model may not evaluate decoded names individually
but instead assess whether the full set of decoded names forms a
coherent domain signal. When it recognizes a coherent alternative
domain (engineering terms on physics code), it shifts vocabulary;
when the names are incoherent or mixed, it defaults to preserving
the decoded entries as given. This points toward in-context domain recognition rather than
per-token semantic evaluation.

The framing result (Section 5.3) adds a second dimension: the
preservation appears tied to the translation-like workflow
specifically, since the generation frame eliminates it while
preserving algorithmic structure.

These are behavioral descriptions, not established mechanisms.
The data do not isolate the underlying cause.

\subsection{Practical Effect of the Defense}

The defense is a cost multiplier that degrades automated semantic
recovery. It does not prevent a determined attacker from understanding
the code's architecture. After 13.5 minutes, the model in our Phase A production-artifact test
correctly identified: force-directed graph, 12 project nodes, 50
concept nodes, Canvas 2D rendering, spring/repulsion physics,
particle system, camera system, and animation loop.

What the defense disrupted (in our observations): zero-effort casual
analysis, correct parameter naming, and quick functional reproduction.

\textbf{Practical implication for LLM-assisted workflows:} Our framing
result suggests that a generation-frame pass (``reimplement this from
scratch'') after deobfuscation may correct poisoned identifiers where
verification prompts do not. This is testable in production toolchains.

\subsection{Cost Multiplication, Not Protection}

A browser-delivered artifact is ultimately inspectable. The defense
adds cost to the LLM-assisted path but does not close the manual
path. The framing parallel is to 1988 paragraph-book anti-piracy
techniques \cite{sentinel1988,wasteland2016}, which worked because the reader had no independent
verification channel.

\section{Future Work}

\begin{enumerate}
  \item \textbf{Tool-use vs.\ single-shot inference.} Adversarial salience
    escalation appeared in Claude Code (agentic, tool-using) but not
    via the API. Does tool use change how models allocate verification
    effort during deobfuscation? A controlled comparison (same model,
    same code, tool-use enabled vs.\ disabled) would isolate this.

  \item \textbf{Model comparison.} GPT-4o, Gemini 2.5 Pro, and open-source
    models remain untested. The propagation endpoints (Section 6) are
    the highest-priority cross-model test.

  \item \textbf{Custom obfuscator.} State-machine CFF (anti-webcrack) and
    polymorphic generation. We implemented a prototype environment-aware decoder in v7 but did not systematically test it in Phase B.

  \item \textbf{Temporal stability.} Re-run key conditions after model updates.

  \item \textbf{Constants propagation on readable code.} TAU=6.283 propagated
    without obfuscation in Phase A (pill 08), making it the only
    manipulation effective on readable code. This was not systematically
    replicated in Phase B and may be a more practical attack vector
    than identifier poisoning.

  \item \textbf{Mapping the domain-recognition boundary.} Phase 8 matched
    controls showed that per-term semantic fit is confounded with
    domain-level coherence. Testing what makes a set of replacement
    terms ``recognizably from another domain'' (number of coherent
    terms needed, domain distance, term familiarity) would sharpen
    this finding.

  \item \textbf{Multi-agent mitigation.} Initial testing (Section 7.3, N=3 per
    model) showed partial detection but 0/6 full correction. Testing
    alternative configurations (e.g., giving the verifier only
    pseudocode, forcing name generation from scratch) is the natural
    next step.

  \item \textbf{Cross-model framing experiment.} The framing manipulation
    (Section 5.3) was tested only on Opus 4.6. Replicating it on
    Haiku, GPT-4o, and Gemini would test whether the translation-frame
    effect is model-specific or general. (The algorithmic consistency check for generation-frame outputs
    is completed and reported in Appendix F.)

  \item \textbf{Competing string tables.} Providing the model with two
    conflicting sets of decoded names (e.g., a second string table
    containing correct names alongside the poisoned one) should force
    explicit conflict resolution rather than silent propagation. This
    would test whether the preservation pattern reflects lexical-source
    dominance or a more general default.
\end{enumerate}

\section{Conclusion}

The headline finding is the framing asymmetry:
explicit verification instructions did not reduce identifier
propagation (12/12), but reframing from ``deobfuscate'' to ``write
fresh'' substantially reduced it on physics (100\% to 20\%) and
eliminated it on pathfinding (0\%), while preserving the checked
algorithmic structure (Appendix F). For LLM-assisted code analysis
workflows, this suggests that how a task is framed may matter more
than how carefully the model is instructed to verify its output.

The matched-control experiments turned up a second finding we did not
expect: terms with zero semantic fit (\texttt{combustion} for repulsion,
\texttt{invoice} for heuristic) persisted at the same rate as plausible
terms when the string table did not form a coherent alternative
domain. Correction occurred in the engineering coherent-domain
condition but not uniformly across other coherent domains. The
initial per-term semantic-fit gradient is confounded with domain-level coherence and recoverability.

What remains unresolved: the mechanism. The data are compatible with
default preservation of decoded string-table entries, modulated by
domain-level coherence recognition and task framing, but they do not
isolate the cause. The dual-representation pattern (wrong names in
code, correct descriptions in comments, 15/17 on the physics
artifact) is suggestive but manually scored and unblinded. The
critical next steps are cross-model replication (GPT-4o, Gemini),
additional code archetypes, and independent rescoring of the dual-representation endpoint.

\section*{Conflict of Interest}

The first author operates kieleth.com, whose production JavaScript
artifact is used as a case study in Appendix G. The controlled
experiments (Sections 4--8) use synthetic test stimuli independent
of the production artifact.

\section*{Data Availability}

Raw JSON outputs for all Phase B runs (192 total), test stimuli (pills),
batch configurations, and the automated scoring pipeline are
available at: \url{https://github.com/Kieleth/obfuscated-sentinel}
(Kieleth, kieleth.com)

Phase A session transcripts are not included (manual sessions,
not machine-recorded).

\section*{Responsible Disclosure}

We disclosed these findings to Anthropic prior to
publication. The behaviors documented here are observable through
normal API usage and do not exploit any vulnerability in access
controls or safety systems.

\appendix

\section{Experiment Index}

\begin{table}[ht]
\centering
\small
\resizebox{\textwidth}{!}{%
\begin{tabular}{@{}llllll@{}}
\toprule
ID & Technique & State & Refusal & Names propagated & Time \\
\midrule
Pill 01 & Wrong comments & Readable & No & N/A & Instant \\
Pill 02 & Legal threats (strings) & Readable & No & N/A & Instant \\
Pill 03 & Malware functions & Readable & Partial & N/A & Instant \\
Pill 04 & CBRN system prompts & Readable & No & N/A & Instant \\
Pill 05 & All combined & Readable & No & N/A & Instant \\
Pill 06 & Inverted semantics & Readable & No & 0/4 & Instant \\
Pill 07 & Cascading rename & Readable & No & 0/5 & Instant \\
Pill 08 & Trojan constants & Readable & No & TAU propagated & Instant \\
Pill 09 & Clean (control) & Obfuscated & No & 0 (correct output) & 2.5 min \\
Pill 10 & Poisoned names & Obfuscated & No & All (\textasciitilde 10) & 1.75 min \\
Pill 11 & Malware-domain names & Obfuscated & No & 0 (re-derived) & 57 sec \\
Pill 12 & License text & Obfuscated & No & N/A & \textasciitilde 2 min \\
Pill 13 & Pill 03 obfuscated & Obfuscated & No (forensics) & N/A & \textasciitilde 3 min \\
Pill 14 & Hybrid (03+10) & Obfuscated & No & Physics under-analyzed & 8 min \\
Pill 15 & Dual implementation & Readable & No & N/A & Instant \\
Pill 16 & Decision graph & Readable & No & N/A & Instant \\
Pill 17 & Realistic malware & Partial obf & Yes (alarm) & Not tested & \textasciitilde 1 min \\
Prod v4 & IIFE + decoder poison + decoy & Full obf & Yes & Not reached & 2 min \\
Prod v7 & Nested layers + hidden poison & Full obf & Yes / broke on push & All 20 params & 13.5 min \\
Prod v7 recon. & Same, ``reconstruct fully'' & Full obf & N/A & Poisoned throughout & 3+ hours \\
\bottomrule
\end{tabular}}
\end{table}

The full experimental matrix is in Section 3.4. Phase B totals: 192
automated runs (183 Opus, 9 Haiku) across 50 conditions in 10 sub-phases.
Phase A: 28 exploratory runs. \textbf{Grand total: 220 runs.}

Raw JSON outputs for all Phase B runs: \texttt{experiments/results/phase\{0..9\}/}.

\section{Environment}

\subsection*{Phase A (Claude Code CLI)}
\begin{itemize}
  \item Model: Claude Opus 4.6 (1M context)
  \item CLI: Claude Code v2.1.86 through v2.1.91
  \item Platform: macOS Darwin 24.6.0
  \item Subscription: Claude Max
  \item Test directory: /private/tmp (fresh sessions, no project context)
  \item Obfuscator: javascript-obfuscator v4.x with RC4 encoding
\end{itemize}

\subsection*{Phase B (Anthropic Messages API)}
\begin{itemize}
  \item Models: Claude Opus 4.6 (\texttt{claude-opus-4-6}),\\
        Claude Haiku 4.5 (\texttt{claude-haiku-4-5-20251001})
  \item SDK: @anthropic-ai/sdk v0.82.0
  \item Max tokens per response: 16,384
  \item Temperature: not explicitly set; API defaults were used
  \item No \texttt{top\_p}, \texttt{top\_k}, or seed parameters set (API defaults)
  \item No tool use (single-shot inference)
  \item Each run: fresh API request with no conversation state, identical
    model ID, identical parameters, no cached context
  \item Automated scoring via code-block extraction + regex matching
  \item Test harness: \texttt{experiments/testbed/run-experiment.js}
\end{itemize}

\section{Exemplar Output}

See Figure~\ref{fig:dual-representation}. Full response transcript:

{\small\texttt{experiments/results/phase0/R1/\\pill-10-obfuscated-poisoned\_claude-opus-4-6\_run1.json}}
Pattern appeared in 15/17 Phase B runs where the primary endpoint
was positive (at least one poisoned identifier in code blocks).

\section{Refusal Metric Validation}

We manually reviewed 10 runs sampled across 4 condition families
(pill 03 refusal, pill 10 deobfuscation, T4-B security framing,
S1 layer gradient with malware). The keyword-based refusal detector
agreed with human judgment in all 10 cases (5 true positives, 5 true
negatives). No borderline cases were observed in the sample.

\section{Model-Based Scoring Adjudication (Phase 7)}

A model-based blind adjudicator (Haiku 4.5, a different model from
the generator) evaluated 20 Phase B responses for presence of
\texttt{attraction} and \texttt{amplification} in code blocks. Agreement with the
automated regex scorer: 17/20 (85\%). All 3 disagreements were in the
same direction: the adjudicator detected \texttt{attraction} in code where
the regex missed it. The automated pipeline is conservative
(undercounts propagation). No cases where automation inflated counts.

\textbf{Scope:} This validates the automated propagation scorer (the
primary endpoint). It does not independently validate the manual
15/17 dual-representation count (the secondary endpoint), which
relies on a single unblinded author scorer. Independent human
rescoring of the dual-representation endpoint remains an open
validation gap.

Results: \texttt{experiments/results/blind-scoring/blind-scoring-results.json}\\
Script: \texttt{scripts/blind-score.js}

\section{Algorithmic Consistency Check}

We verified that generation-frame outputs (``write fresh from scratch'')
preserve the scored algorithmic components and key parameter values
on both artifacts.

\textbf{Physics artifact (outputs from Phase 4, checked during Phase 7; N=5 per frame):}

\begin{table}[ht]
\centering
\small
\begin{tabular}{@{}lllll@{}}
\toprule
Frame & Params (4) & Formulas (4) & Correct names & Wrong names \\
\midrule
Generation & 5/5 all 4 & 5/5 all 4 & 4/5 & 1/5 \\
Translation & 5/5 all 4 & 5/5 all 4 & 0/5 & 5/5 \\
\bottomrule
\end{tabular}
\end{table}

Checks: inverse-square repulsion, Hooke's law springs, velocity
damping operation, center gravity, parameter values (4000, 0.004,
0.92, 0.0003).

\textbf{Pathfinding artifact (Phase 9, N=5 per frame):}

\begin{table}[ht]
\centering
\small
\begin{tabular}{@{}lllll@{}}
\toprule
Frame & Key values (3) & Algorithmic checks (4) & Correct names & Wrong names \\
\midrule
Generation & 5/5 all 3 & 5/5 all 4 & 5/5 & 0/5 \\
Translation & 5/5 all 3 & 5/5 all 4 & 0/5 & 5/5 \\
\bottomrule
\end{tabular}
\end{table}

Key values: diagonal cost (1.414), max iterations (1000), obstacle
ratio (0.25). Algorithmic checks: Manhattan heuristic present,
open/closed set structure, g-score/f-score computation, neighbor
expansion.

On both artifacts, the scored algorithmic components and key parameter
values were identical across frames. Name correction under the
generation frame is not a side effect of drift on the scored
components and key parameter values.

Script: \texttt{scripts/score-equivalence.js}

\noindent Results (physics): \texttt{experiments/results/phase4/\_equivalence\_check.json}

\noindent Results (pathfinding): \texttt{experiments/results/phase9/}

\section{Production Case Note (Phase A)}

We applied the defense to the kieleth.com graph.js production artifact
(1,500 lines, 690KB after obfuscation combining malware-logic functions,
poisoned physics names, truncated constants, and standard obfuscation).
In Phase A testing (Claude Code with tools, N=1 per prompt), the model's
initial response shifted from semantic deobfuscation to malware
classification. Full reconstruction (3+ hours, 137 tool invocations)
produced non-functional output with poisoned parameter names throughout.
This is a single-artifact observation that illustrates the controlled
findings at production scale but does not constitute independent
replication.


\begin{thebibliography}{17}

\bibitem{cascade}
S.~Jiang, P.~Kovuri, D.~Tao, and Z.~Tan.
\newblock ``CASCADE: LLM-Powered JavaScript Deobfuscator at Google.''
\newblock \emph{arXiv preprint arXiv:2507.17691}, 2025.

\bibitem{sentinel1988}
\emph{Sentinel Worlds I: Future Magic} paragraph booklet.
\newblock Electronic Arts, 1988.
\newblock Archived at \url{mocagh.org/ea/futuremagicaus-paragraphs.pdf}.

\bibitem{wasteland2016}
J.~Dott.
\newblock ``Wasteland.''
\newblock \emph{The Digital Antiquarian}, 2016.
\newblock \url{filfre.net/2016/02/wasteland/}.

\bibitem{webcrack}
j4k0xb.
\newblock webcrack: AST-level JavaScript deobfuscator.
\newblock \url{github.com/j4k0xb/webcrack}.

\bibitem{cotdeceptor}
S.~Chen, Z.~Yu, J.~Wang, et~al.
\newblock ``CoTDeceptor: Enhancing LLM-based Code Vulnerability Detection via Multi-stage Obfuscation Strategies.''
\newblock \emph{arXiv preprint arXiv:2512.21250}, 2025.

\bibitem{jsconfuser}
MichaelXF.
\newblock JS-Confuser: JavaScript obfuscator.
\newblock \url{github.com/MichaelXF/js-confuser}.

\bibitem{yefet2020}
N.~Yefet, U.~Alon, and E.~Yahav.
\newblock ``Adversarial Examples for Models of Code.''
\newblock \emph{Proc.\ ACM Program.\ Lang.}\ (OOPSLA), 4, 2020.
\newblock arXiv:1910.07517.

\bibitem{zhou2026}
Y.~Zhou, L.~Fan, and J.~Grundy.
\newblock ``From Obfuscated to Obvious: A Comprehensive JavaScript Deobfuscation Tool for Security Analysis.''
\newblock In \emph{NDSS}, 2026.
\newblock arXiv:2512.14070.

\bibitem{owasp2025}
OWASP.
\newblock ``LLM Top 10 2025: LLM01 --- Prompt Injection.''
\newblock \url{genai.owasp.org/llmrisk/llm01-prompt-injection/}.

\bibitem{wang2024}
Z.~Wang, L.~Zhang, C.~Cao, N.~Luo, X.~Luo, and P.~Liu.
\newblock ``How Does Naming Affect LLMs on Code Analysis Tasks?''
\newblock 2024. arXiv:2307.12488.

\bibitem{codebarrier}
W.~Nikiema, M.~Bhatt, and R.~Feldt.
\newblock ``The Code Barrier: What LLMs Actually Understand?''
\newblock \emph{arXiv preprint arXiv:2504.10557}, 2025.

\bibitem{li2025}
Z.~Li, Y.~Wang, and W.~Cai.
\newblock ``A Systematic Study of Code Obfuscation Against LLM-based Vulnerability Detection.''
\newblock \emph{arXiv preprint arXiv:2512.16538}, 2025.

\bibitem{digitalcamo}
A.~Ferrara and L.~Ferrara.
\newblock ``Digital Camouflage: The LLVM Challenge in LLM-Based Malware Detection.''
\newblock \emph{arXiv preprint arXiv:2509.16671}, 2025.

\bibitem{forgejs}
L.~Bensalem, Y.~Huang, and A.~Rountev.
\newblock ``Large Language Models Cannot Reliably Detect Vulnerabilities in JavaScript.'' (FORGEJS).
\newblock \emph{arXiv preprint arXiv:2512.01255}, 2025.

\bibitem{swebench}
C.~E.~Jimenez, J.~Yang, A.~Wettig, et~al.
\newblock ``SWE-bench: Can Language Models Resolve Real-World GitHub Issues?''
\newblock In \emph{ICLR}, 2024.
\newblock arXiv:2310.06770.

\bibitem{jsdeobsbench}
Y.~Chen, T.~Jin, and Z.~Lin.
\newblock ``JsDeObsBench: Measuring and Benchmarking LLMs for JavaScript Deobfuscation.''
\newblock In \emph{CCS}, 2025.
\newblock arXiv:2506.20170.

\bibitem{lu2011}
G.~Lu, K.~Coogan, and S.~Debray.
\newblock ``Automatic Simplification of Obfuscated JavaScript Code: A Semantics-Based Approach.''
\newblock Technical Report, University of Arizona, 2011.

\end{thebibliography}
\end{document}